\documentclass[a4paper,11pt]{article}
\pdfoutput=1 

\usepackage{jheppub} 

\usepackage[T1]{fontenc} 

\usepackage{xcolor}

\usepackage{amsmath}

\title{\boldmath Determination of the first-generation quark couplings at the $Z$-pole}

\author[a,b,1]{K. M\k{e}ka{\l}a,\note{Corresponding author.}}
\author[c]{D. Jeans,}
\author[b]{J. Reuter,}
\author[d]{J. Tian,}
\author[a]{and A.F. \.Zarnecki}

\affiliation[a]{Faculty of Physics, University of Warsaw, Pasteura 5, 02-093 Warsaw, Poland}
\affiliation[b]{Deutsches Elektronen-Synchrotron DESY, Notkestr. 85, 22607 Hamburg, Germany}
\affiliation[c]{KEK, 1-1 Oho Tsukuba, Ibaraki 305-0801, Japan}
\affiliation[d]{International Center for Elementary Particle Physics ICEPP, University of Tokyo, 7-3-1 Hongo, Bunkyo-ku, Tokyo 113-0033, Japan}

\emailAdd{k.mekala@uw.edu.pl}
\emailAdd{daniel.jeans@kek.jp}
\emailAdd{juergen.reuter@desy.de}
\emailAdd{tian@icepp.s.u-tokyo.ac.jp}
\emailAdd{filip.zarnecki@fuw.edu.pl}

\abstract{Electroweak Precision Measurements are stringent tests of the Standard Model and sensitive probes to New Physics. Accurate studies of the $Z$-boson couplings to the first-generation quarks, which are currently constrained from LEP data to a few percent, could reveal potential discrepancies from the theory predictions. Future $e^+e^-$ colliders running at the $Z$-pole would be an excellent tool for an analysis based on a comparison of radiative and non-radiative $Z$ boson decays. In this paper, we present a method to extract the values of the $Z$ couplings to light quarks and discuss the uncertainty of the measurement, including contributions from various systematic effects. We show that systematic uncertainty in the heavy-flavour tagging performance is the key factor in the analysis and reducing it to a sub-permille level might be crucial to fully profit from the high luminosity of future $e^+e^-$ machines. The measurement could improve the LEP results by at least an order of magnitude.
}

\begin{document} 
\maketitle
\flushbottom

\section{Motivation}
The Standard Model (SM) of particle physics is the best theory to describe fundamental interactions at excellent precision. Although several experiments, including cosmological observations, point to the existence of exotic components of matter building the Universe, to date no single particle has been found Beyond the Standard Model (BSM). To reveal the unknown, the particle physics community agrees on the necessity of exploring new directions, after completing the Large Hadron Collider (LHC) programme, by building an $e^+ e^-$ Higgs factory, as expressed in the 2020 Update of the European Strategy for Particle Physics~\cite{EuropeanStrategyGroup:2020pow} and the P5 Report in the US~\cite{P5:2023wyd}. Not only would such a collider improve the measurement of the properties of the Higgs boson, but it would also allow for direct searches of New Physics, as well as deliver the most precise constraints on the parameters in the electroweak sector. The latter is perceived as the main reason behind starting the operation of the new facility at the $Z$-pole, collecting unprecedented statistics of $Z$ bosons in a clean collision environment. The $Z$-pole runs are currently considered for most facilities: CEPC~\cite{CEPCStudyGroup:2018ghi,CEPCStudyGroup:2023quu}, FCC-ee~\cite{FCC:2018byv}, ILC~\cite{Behnke:2013xla,ILC:2013jhg,Behnke:2013lya},  and LCF~\cite{LinearColliderVision:2025hlt, LinearCollider:2025lya}, differing in achievable luminosities ($10^2 -10^5$~fb$^{-1}$).
The measurement discussed in this paper is not sensitive to beam polarisation assumed for linear machines, and we will not consider it in the following.

Among others, the huge collected statistics of $Z$ bosons could be used to measure its electroweak couplings to quarks. The LEP experiments achieved per-mille precision in constraining the partial width of the $Z$-boson decaying into $b\bar{b}$ pairs and percent precision for $c\bar{c}$ pairs~\cite{ParticleDataGroup:2024cfk}. Improvements in detector technologies and heavy-quark tagging algorithms\footnote{Throughout the paper, we will refer to the $u$ and $d$ quarks as ``light'' and to the others, including the $s$ quark, as ``heavy''.}, are expected to significantly improve these results at future colliders (see e.g. \cite{Bilokin:2017lco, Fuster:2021ekh, Irles:2023ojs,Tagami:2024gtc,Liang:2023wpt,Zhu:2023xpk,Blekman:2024wyf,Albert:2022mpk}). However, due to the lack of efficient tagging algorithms, direct measurements for light quarks are challenging, and an alternative approach is required. Since the probability of photon emission is related to the electric charge of the particle, up-type quarks emit photons with much higher probability compared to down-type quarks. Conversely, down-type quarks dominate the non-radiative sample due to the larger value of the SM electroweak coupling. These observations can be leveraged to study the hadronic decays of the $Z$ boson by simultaneously examining both radiative and non-radiative signatures, thereby disentangling their dependence on the value of the couplings. Similar measurements were taken at LEP~\cite{DELPHI:1991utk, L3:1992kcg, L3:1992ukp, DELPHI:1995okj, OPAL:2003gdu} and yielded results of up to 3\% uncertainty for $d$ and 6\% for $u$ quarks. Our analysis aims to extend this work towards future colliders, taking into account modern experimental developments.

In this paper, we introduce the proposed method and examine its applicability to future  $e^+ e^-$ colliders operating at the $Z$ pole. The conceptual framework for the measurement is outlined in Section \ref{sec:idea}. Section \ref{sec:simulation} details our event-generation procedure and theoretical modelling. The subsequent Section \ref{sec:analysis} presents the analysis framework and the obtained results. Finally, we provide our conclusions and outlook in Section \ref{sec:conclusions}. Additional information on the employed statistical methods is provided in the Appendix \ref{sec:app_stat}.
\section{Outline of the measurement}
\label{sec:idea}
The concept of the measurement relies on the fact that up- and down-type quarks differ in electric charge and thus, their electromagnetic couplings are distinguishable. The coupling strength of the $Z$ boson to a given fermion $f$ is conventionally defined as
\begin{equation}
    c_f = v^2_f + a^2_f,
\end{equation}
where $v_f$ ($a_f$) is the vector (axial) coupling. In the SM, the couplings are expressed in terms of the third component of the fermion weak isospin ($I_{3,f}$) and its charge ($Q_f$):
\begin{equation}
    v_f = 2I_{3,f}-4Q_f \sin^2{\theta_W}, \quad \quad a_f = 2I_{3,f},
\end{equation}
where $\theta_W$ stands for the weak mixing angle.
The total width of the $Z$ boson to hadrons, $\Gamma_{had}$, is given at fixed order in $\alpha_s$ by~\cite{L3:1992kcg,OPAL:2003gdu,ParticleDataGroup:2024cfk}
\begin{equation}
    \Gamma_{had} = N_c \frac{G_\mu M^3_Z}{24\pi \sqrt{2}}\left(1 + \frac{\alpha_s}{\pi}+ \mathcal{O}(\alpha_s^2)\right) \sum_{q = u,d,s,c,b} c_q,
\end{equation}
where $N_c$ is the number of colours, $G_\mu$ is the Fermi constant, $M_Z$ is the mass of the $Z$ boson and $c_q$ is the coupling to a given quark\footnote{From now on, we will always assume that the sum runs over five quark flavours.}, respectively. The total width to radiative hadronic decays, $\Gamma_{had+\gamma}$, can be expressed at leading QED order as
\begin{equation}
    \Gamma_{had+\gamma} = N_c \frac{G_\mu M^3_Z}{24\pi \sqrt{2}} f(y_{cut})\frac{\alpha}{2\pi} \sum_{q} c_q Q_q^2 ,
\end{equation}
where $f(y_{cut})$ is an acceptance factor depending on a parameter $y_{cut}$ incorporating the isolation criteria for photons, $\alpha$ is the electromagnetic coupling constant and $Q_q$  is the electric charge of the given quark. Since the electric charges of $u$-type and $d$-type quarks are different, the expressions for the radiative and the total hadronic widths include different coupling combinations, and the dependence on the couplings $c_q$ can potentially be resolved.

Similarly, the hadronic cross section on the $Z$ pole, $e^+e^- \to Z \to q\bar{q}$, $q = u, d, s, c, b$, can be expressed as
\begin{equation}
    \sigma_{had} = \sum_{q} \sigma_q  = \mathcal{C}_1 \cdot \sum_{q} c_q, 
\end{equation}
where $\mathcal{C}_1$ is a constant. The radiative hadronic cross section with exactly one photon identified in the final state (the one-photon inclusive cross section) can be parametrised as
\begin{equation}
       \sigma_{had+\gamma} = \mathcal{C}_2 \cdot \sum_{q} c_q Q_q^2,
\end{equation}
where $\mathcal{C}_2$ is another constant (for a given set of cuts and isolation criteria for photons). The values of $\mathcal{C}_1$ and $\mathcal{C}_2$ can be estimated from theoretical calculations and simulations.

In the simplest case, it suffices to identify light-quark events by excluding contributions from heavy-quark events through flavour tagging algorithms. Fitting the experimental data to theory predictions allows for the extraction of $c_u$ and $c_d$ using the well-established maximum likelihood method. More details on the generalised formalism including the impact of systematic uncertainties are given in Appendix \ref{sec:app_stat}.

\section{Simulation of events}
\label{sec:simulation}
High accuracy of the measurement can only be achieved if experimental data is compared with realistic and precise Monte Carlo simulations. Although the process of $e^+e^- \to q\bar{q}$ is one of the simplest electroweak processes and thus often perceived as a benchmark point for event generators (cf. e.g.~\cite{Robson:2920434}), the reconstruction of isolated photons poses several challenges. A proper description of the process necessitates the full Matrix-Element (ME) generation of the measurable photons. However, soft and collinear photons must be technically generated using modelling methods, such as parton showers. Additionally, due to the background originating from hadron decays, hadronisation effects must also be incorporated. Thus, as a recipe, one can generate data samples using fixed-order ME calculations, with exclusive emissions of hard photons, and match them with initial-state radiation (ISR) structure functions and final-state radiation (FSR) showers, accounting for collinear and soft emissions, and hadronisation models.

A similar issue was investigated in \cite{Kalinowski:2020lhp} where dark-matter production detected in the mono-photon signature at lepton colliders was considered. A matching procedure for simulating photons using both ME calculations (for hard emissions) and ISR structure function (for soft emissions) was developed for \textsc{Whizard}~\cite{Kilian_2011,moretti2001omega}, a Monte Carlo generator incorporating many features suitable for future lepton colliders, including beam polarisation, beamstrahlung and ISR spectra. \textsc{Whizard} uses the colour-flow formalism for QCD partons~\cite{Kilian:2012pz}. For the purpose of this study, it is sufficient to consider the leading-order contributions in QCD and electroweak interactions, though \textsc{Whizard} is capable of automated NLO QCD+EW corrections~\cite{Bredt:2022dmm}.
As the processes simulated here are relatively simple, we do not have to make use of the extensive parallelisation capabilities of \textsc{Whizard}~\cite{Brass:2018xbv,ChokoufeNejad:2014skp}.
The procedure of \cite{Kalinowski:2020lhp} was validated against \textsc{KKMC}~\cite{Jadach:1999vf,Jadach:2013aha}. We follow and extend this approach to include effects occurring for hadronic final states. Thus, parton-level events are generated with \textsc{Whizard}, and \textsc{Pythia\,6}~\cite{Sjostrand:2006za} is employed to simulate parton showers and hadronisation. A standard cut on the invariant mass of the parton-level hadronic system to be above 10 GeV was applied. It was verified that the obtained results do not differ qualitatively when using  \textsc{Pythia\,8}~\cite{Sjostrand:2014zea}.

The matching in \cite{Kalinowski:2020lhp} is based on two variables, $q_-$ and $q_+$, defined for each photon as:
\begin{align*}
    q_- = \sqrt{4E_0 E_\gamma}\sin \frac{\theta_\gamma}{2},\\
    q_+ = \sqrt{4E_0 E_\gamma}\cos \frac{\theta_\gamma}{2},
\end{align*}
where $E_0$ is the nominal beam energy, $E_\gamma$ is the energy of the emitted photon and $\theta_\gamma$ is its emission angle. The variable $q_-$ ($q_+$) corresponds to the virtuality of an electron (positron) after a single photon emission. According to the procedure, all the photons with $q_\pm > q_{min}$  and $E_\gamma > E_{min}$ are generated in the fixed-order calculation picture, while all the \textit{softer} emissions are modeled via the built-in ISR structure function handler for the ISR photons in \textsc{Whizard}. 

The procedure of \cite{Kalinowski:2020lhp} was developed for chargeless particles in the final state (electrically neutral and colourless). Hadronic decays of the $Z$ bosons require an extension of the matching to include final-state QCD and QED showers. Thus, to separate \textit{hard} and \textit{soft} final-state radiation regimes, we use an additional criterion based on the invariant mass of the photon-quark pairs, $M_{\gamma\, q_{1,2}}$. A \textit{hard} photon is defined as one fulfilling all the following criteria:
\begin{itemize}\setlength\itemsep{-0.2em}
    \item $q_\pm > 0.5$ GeV,
    \item $E_\gamma > 0.5$ GeV,
    \item $M_{\gamma\, q_{1,2}} > 1$ GeV,
\end{itemize}
and is generated using fixed-order ME calculations. All remaining \textit{soft} photons (those not meeting at least one of those criteria) are simulated via the internal structure function handler for ISR photons in \textsc{Whizard} and the \textsc{Pythia} parton shower for FSR. Events with at least one ISR or FSR photon passing the \textit{hard} photon selection are rejected to avoid double counting. 

We generated samples with 10 million events with 0 ME photons, 10 million events with 1 ME photon and 1 million events with 2 ME photons. 
Higher photon multiplicities have been neglected, as the cross section for the 2-ME-photon sample is already about 30 times smaller than that for the 1-ME-photon sample. The samples were then matched according to the procedure described above.
Table \ref{tab:cs} summarises the cross sections for each flavour for all the \textit{unmatched} samples (with generator-level cuts applied only to ME photons, but not those modelled in showers) and the final matched sample. As expected, the cross section for the matched samples is consistent with the cross sections of the unmatched 0-ME-photon samples.
\begin{table}[h!]
\begin{tabular}{c|c|c|c|c|c|c}
             $\sigma_q$ [fb] & $d$ & $u$ & $s$ & $c$ & $b$ & \textbf{total} \\ \hline
0$\gamma$ unmat.   & 5.46e+06   & 4.28e+06   & 5.46e+06   & 4.28e+06   & 5.47e+06   & 2.49e+07     \\ \hline
1$\gamma$ unmat.   & 3.63e+05   & 4.63e+05   & 3.63e+05   & 4.65e+05   & 3.64e+05   & 2.02e+06     \\ \hline
2$\gamma$ unmat.   & 8.48e+03   & 2.19e+04   & 8.51e+03   & 2.19e+04   & 8.50e+03   & 6.92e+04     \\ \hline
0+1+2$\gamma$ mat. & 5.44e+06   & 4.34e+06   & 5.45e+06   & 4.33e+06   & 5.45e+06   & 2.50e+07     \\
\end{tabular}
\caption{Cross sections for each quark flavour for unmatched samples (generator-level cuts applied only to ME photons) and the final matched sample. Integration uncertainties are below $10^{-3}$.}
\label{tab:cs}
\end{table}

\begin{figure}
    \centering
    \includegraphics[width=0.6\textwidth]{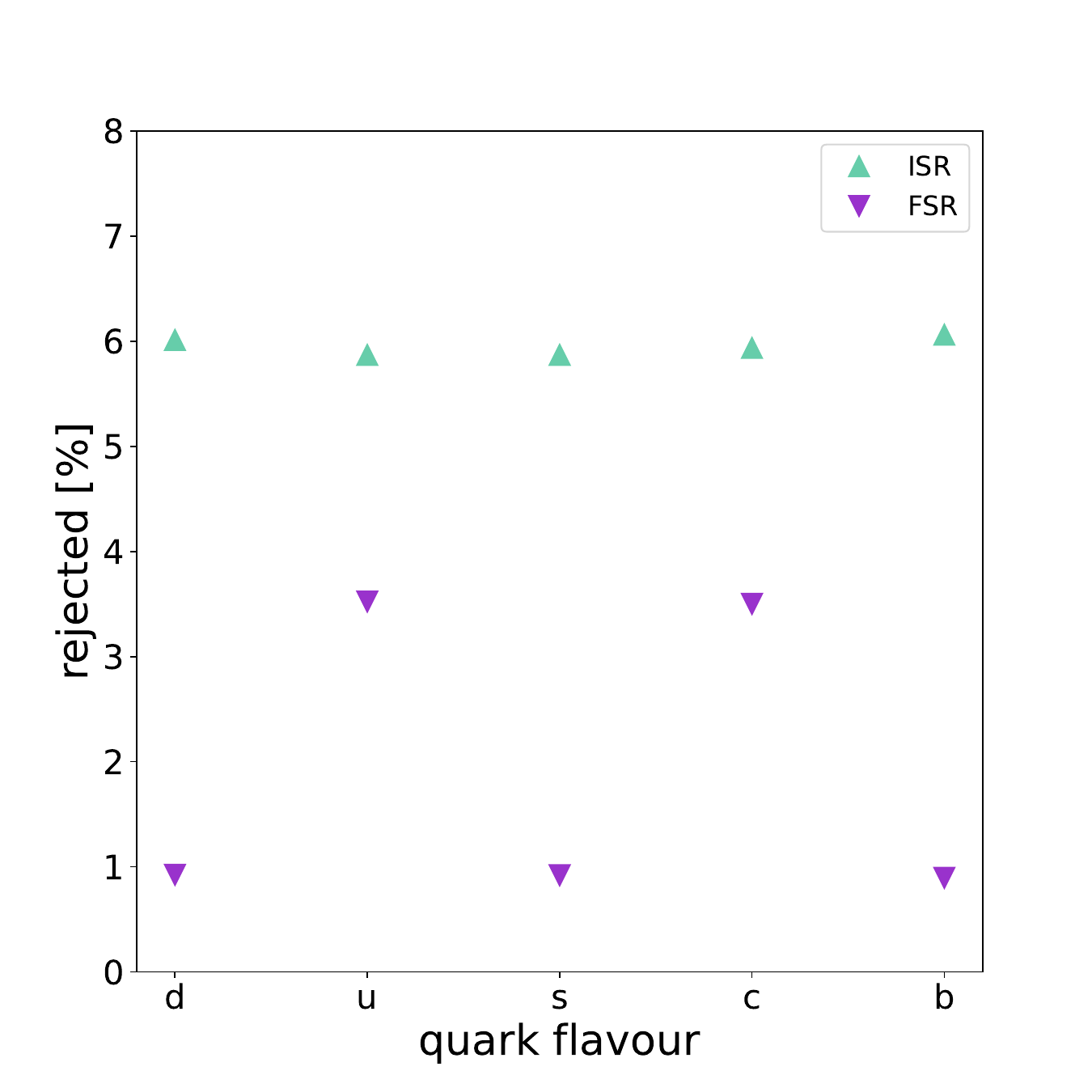}
    \caption{The fraction of events from the sample with no photons generated at the ME level rejected by the matching procedure for different quark flavours (normalised to the cross section per flavour) for ISR (aquamarine upwards) and FSR (purple downwards).}
    \label{fig:rejection}
\end{figure}

Figure~\ref{fig:rejection} shows the fraction of rejected events in the matching procedure for the sample with no photons generated at the matrix-element level for different quark flavours, normalised to the cross section per flavour, for ISR and FSR. As expected, the rejection efficiency for the ISR does not depend on the quark flavour while the rejection efficiency for the FSR distinguishes up- and down-type quarks and differs by a factor of about four between the two cases, which corresponds to the difference in the quark charge squared.

To account for experimental effects, we simulated detector response in \textsc{Delphes 3.5}~\cite{deFavereau:2013fsa}, using the ILCgen cards based on the performance of the ILD and SiD concepts~\cite{ILDConceptGroup:2020sfq, White:2015dxj}. Although originating from linear-collider studies, ILD is currently also considered for circular colliders~\cite{Einhaus:2023npl,Robson:2920434}.
\section{Analysis procedure and results}
\label{sec:analysis}
\subsection{Event reconstruction}
One of the advantages of the measurement at the $Z$-pole is the suppression of the ISR background contribution due to the small phase space for photon emissions. Nevertheless, background photons originating from hadronic decays must be considered in the analysis. Photon reconstruction criteria should be optimised at the experimental level to enhance the contribution from events with the actual hard emissions. We ignore other background channels to the inclusive sample of hadronic $Z$ decays, as their impact is expected to be negligible.

At the ME level, we define the photon isolation parameter as
\begin{equation}
    q^T = E_\gamma \sin \theta^{min}_{\gamma q_i},
\end{equation}
where $E_\gamma$ is the photon energy and $\theta^{min}_{\gamma q_i}$ is the angle between the photon and the closest quark. The variable corresponds to the photon transverse momentum relative to the quark direction. The redefinition of the variable to the detector-level analysis is straightforward: the angle $\theta^{min}_{\gamma q_i}$ is replaced by $\theta^{min}_{\gamma j_i}$, the angle between the photon and the closest jet.

The Durham jet algorithm in the exclusive mode was used in \textsc{Delphes} to cluster all final state objects into two jets.
The photon isolation criteria were not applied in \textsc{Delphes}, so all photons were included in the jet clustering. It allowed us to separate radiative event tagging from the standard reconstruction procedure and treat each photon equally, focusing on the analysis-specific approach\footnote{In this context, the physical interpretation of the $q^T$ variable becomes somewhat vague, as including all photons in jets changes jet kinematics. We will, however, consistently follow our definition which remains comprehensive at the analysis level.}.

\begin{figure}
    \centering
    \includegraphics[width=0.43\textwidth]{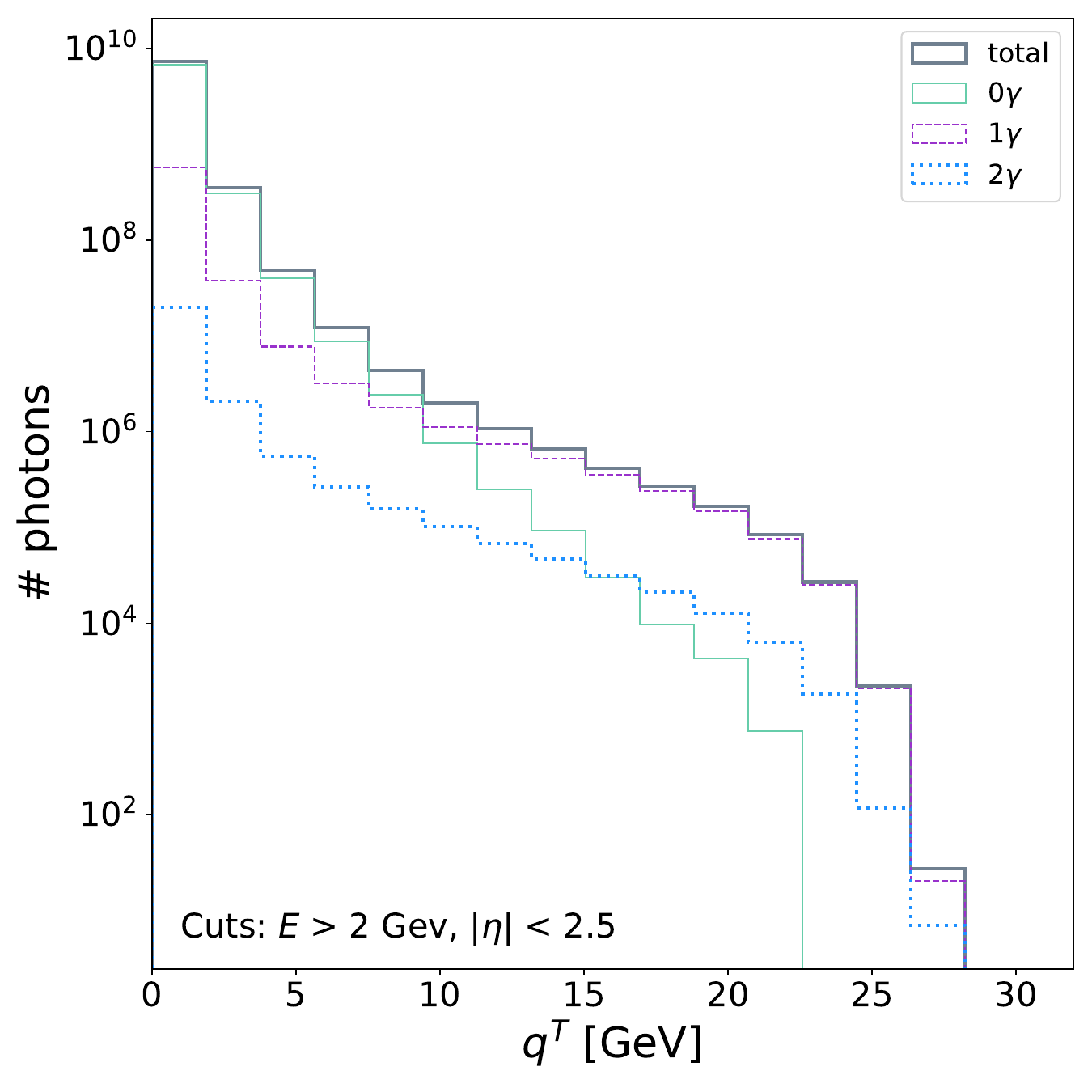}
    \includegraphics[width=0.43\textwidth]{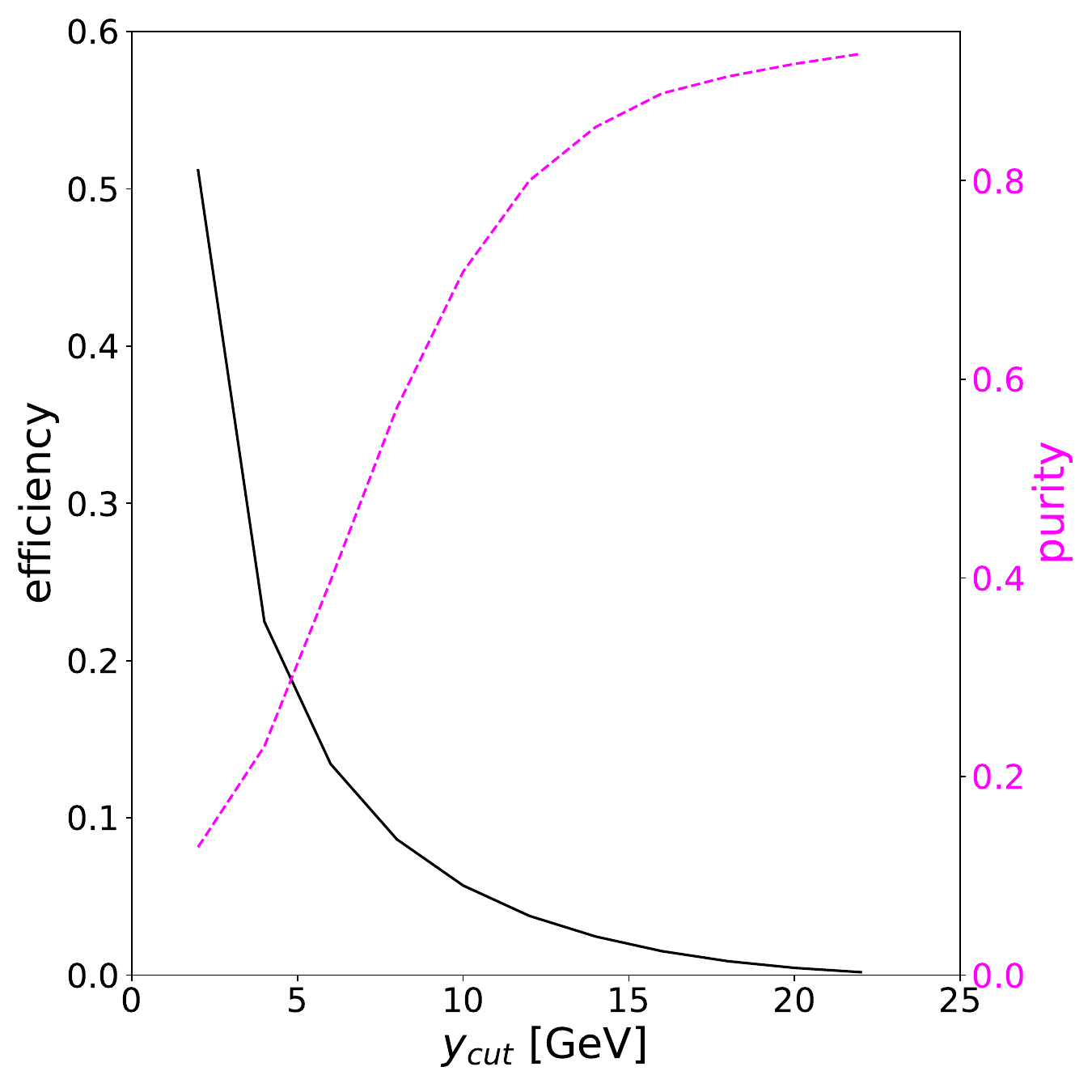}
    
    \caption{\textbf{Left:} the distribution of measurable photons (normalised to an integrated luminosity of 100 fb$^{-1}$) as a function of $q^T$. The thick gray line stands for the sum of all the samples, the solid aquamarine line for the 0-ME-photon sample, the purple dashed one for the 1-ME-photon sample and the blue dotted one -- for the 2-ME-photon sample. Fast detector simulation is included in the results. \textbf{Right:} the signal efficiency (left axis, black solid line) and the signal purity (right axis, pink dashed line) for different values of $y_{cut}$, starting at a minimal value of 2 GeV considered in the study. Note that the distributions are defined at the event level, i.e. they relate to the number of those events in which exactly one photon was tagged.}
    \label{fig:y}
\end{figure}

Figure \ref{fig:y} (left) shows the distribution of $q^T$ for all reconstructed photons at the detector level. The number of events corresponds to a future $e^+e^-$ collider collecting 100 fb$^{-1}$ of unpolarised data and can be easily scaled for other luminosities. Only photons with an energy above 2 GeV and an absolute value of pseudorapidity below 2.5 were assumed to be measured. For low values of $q^T$, the total distribution is dominated by the 0-ME-photon sample (for which the reconstructed photons originate from hadronic decays, with 90\% of the photons coming from $\pi^0$ decays, 6.4\% from $\eta$ decays and 0.9\% coming from $\omega(782)$ decays), while for larger values the 1-ME-photon sample becomes dominant. The transition between the two regions occurs at about 10 GeV. For such values, the 1-ME-photon sample contains only hard photons, i.e. those generated at the ME level and not modelled in showers. Figure \ref{fig:y} (right plot) shows the \textit{signal} efficiency (ratio of the accepted radiative events over all the radiative events) and purity (ratio of the accepted radiative events over all the accepted events) in the sensible range of possible values of $y_{cut}$. For the purpose of this study, we consider 
as signal only those events with exactly one hard tagged photon, i.e. for which $q^T > y_{cut}$. Events with one proper photon generated at the ME level become dominant for a cut value of about 7 GeV, for which about 10\% of events are preserved.

\subsection{2-flavour fit}
In the next step, we used the statistical framework described in Appendix \ref{sec:app_stat}. The quark flavour tagging efficiencies were adapted from \cite{Liang:2023wpt} assuming that both jets are tagged independently and light quarks have exactly the same tagging probabilities\footnote{The latter follows from the fact that the goal of our analysis is to separate the light-quark couplings using photon radiation. The values for the $u$ and $d$ tagging of \cite{Liang:2023wpt} were averaged.}. Each jet was assigned one of the four available jet labels: ``light jet'', ``$s$ jet'', ``$c$ jet'', or ``$b$ jet''.

Firstly, we considered a sample of events with two tagged ``light jets'' only, i.e. we rejected all events in which at least one jet was not classified into the first category. We applied a transverse momentum cut $p^T_{j_i} > 10$ GeV and  a pseudorapidity cut $|\eta_{j_i}| < 4$ on both jets. In this sample, we identified a subsample consisting of events with at least one photon tagged ($E_\gamma > 2$ GeV, $|\eta|<2.5$). The numbers of events in the main inclusive sample (referred to as ``hadronic'' in the following) and its subsample (``radiative'') were used to fit the data and extract $c_d$ and $c_u$ using $\chi^2$ minimisation. In our approach, the radiative subsample has not been removed from the hadronic sample; an alternative approach could be followed, namely removing the radiative events and changing the fit procedure accordingly, but it should not change the final results.

\begin{figure}
    \centering
    \includegraphics[width=0.46\textwidth]{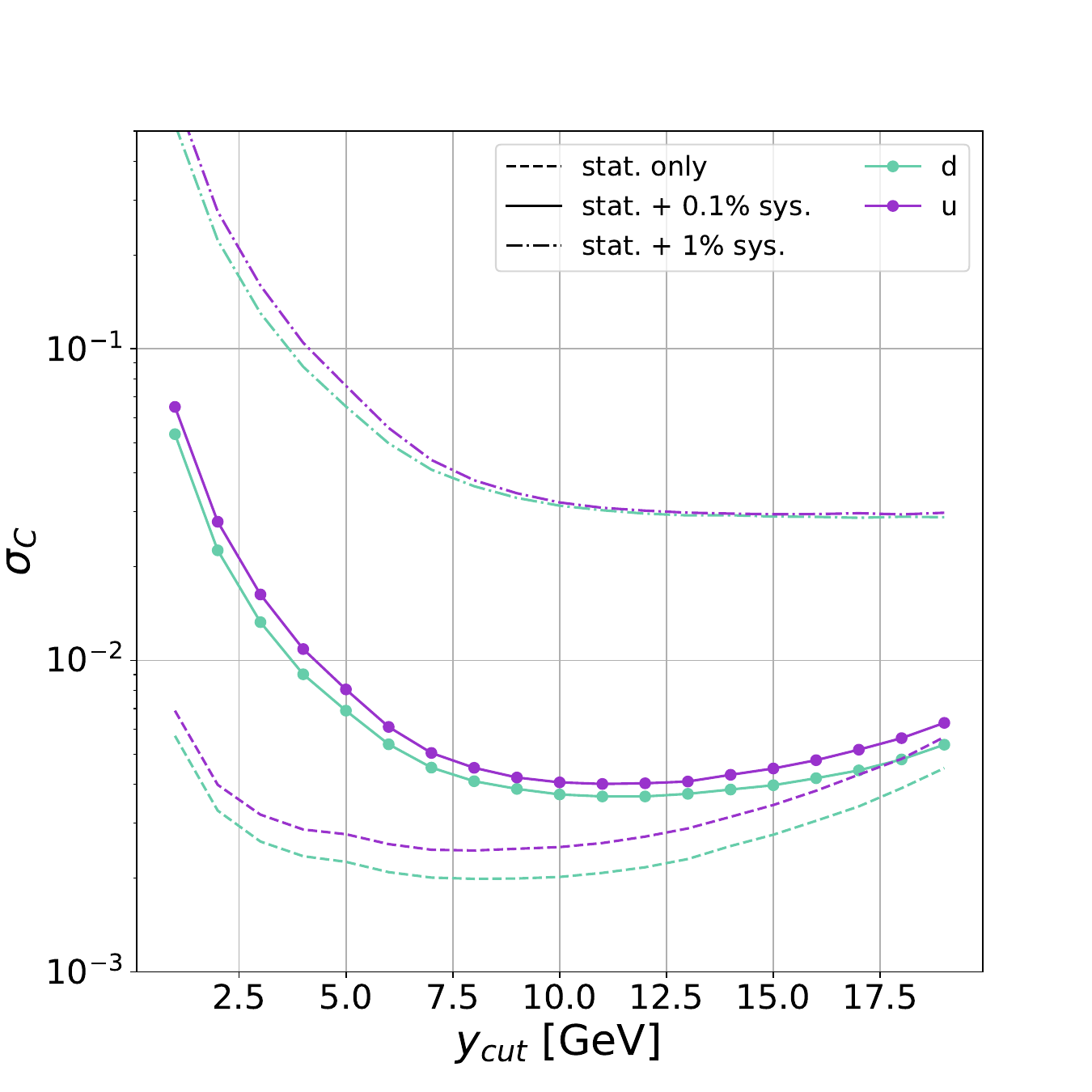}
    \includegraphics[width=0.46\textwidth]{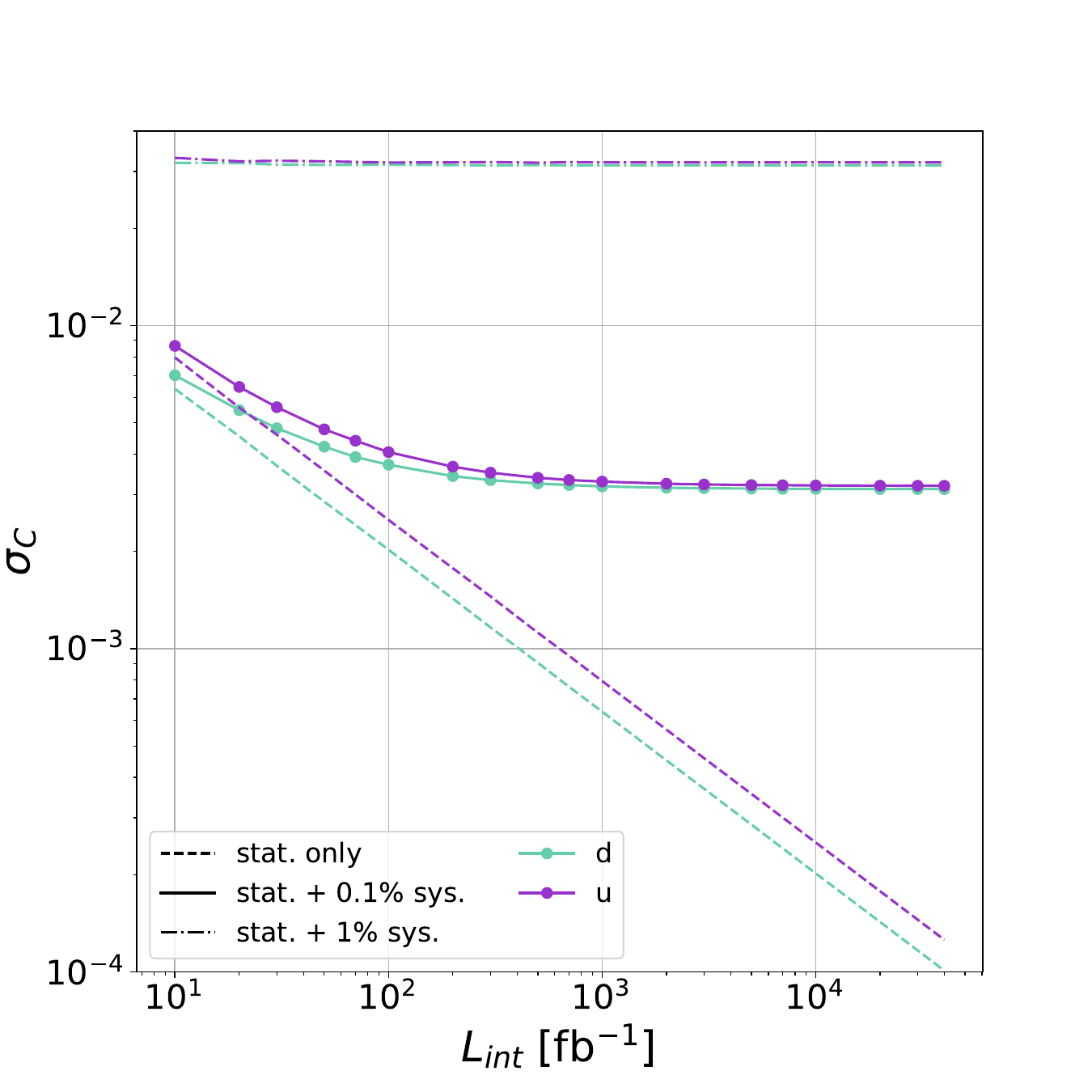}
    
    \caption{The uncertainty of the $d$ (aquamarine) and $u$ (purple) coupling measurement as a function of $y_{cut}$ (left) and the integrated luminosity (right). Dashed lines indicate statistical uncertainty only, solid lines with points -- statistical and systematic uncertainties of 0.1\% combined, dash-dotted lines -- statistical and systematic uncertainties of 1\% combined. The luminosity uncertainty was fixed at 10$^{-4}$. For the left plot, we assume collecting 100 fb$^{-1}$ of data and for the right plot, we set the value of $y_{cut}$ to 10 GeV.}
    \label{fig:2flav}
\end{figure}

Subsequently, we studied the impact of systematic uncertainties, extending the fit procedure as described in Appendix \ref{sec:app_stat}. We assumed that the measurement is affected by the following factors:
\begin{itemize}\setlength\itemsep{-0.2em}
    \item relative uncertainty on the integrated luminosity,
    \item relative uncertainty on the radiative event selection efficiency,
    \item relative uncertainty on the background to the radiative sample due to photons coming from hadronisation,
    \item tagging uncertainties (introduced for all tagging and mis-tagging probabilities used in the fit procedure)\footnote{Setting them as an absolute shift to the elements of the tagging matrix automatically ensures that the tagging uncertainty for the light quarks is larger than for the heavy quarks, as the corresponding elements of the tagging matrix are smaller and an absolute shift of the same magnitude is more significant.}.
\end{itemize}
For simplicity, we initially set uncertainties to the same value for all the contributions, except for the luminosity uncertainty which was always kept at $10^{-4}$ (see e.g. \cite{BozovicJelisavcic:2013lni, Jadach:2018jjo, Dam:2021sdj}). The results are shown in Figure \ref{fig:2flav}, where the relative statistical uncertainties on the $u$ and $d$ quark couplings are compared with the total uncertainties calculated assuming the systematic uncertainty value to be either 0.1\% or 1\%. The left plot presents the dependence of the results on the $y_{cut}$ parameter; as naively estimated from Figure \ref{fig:y}, the optimal value of the parameter corresponds to about 10 GeV, where the FSR-initiated photons start to dominate over photons coming from hadronic decays. The right plot shows the dependence on the integrated luminosity. It is evident that this measurement is systematic-dominated and achieving higher luminosity does not improve the results. In the following, we will discuss potential improvements in the measurement and the impact of particular systematic effects.

\begin{figure}
    \centering
    \includegraphics[width=0.46\textwidth]{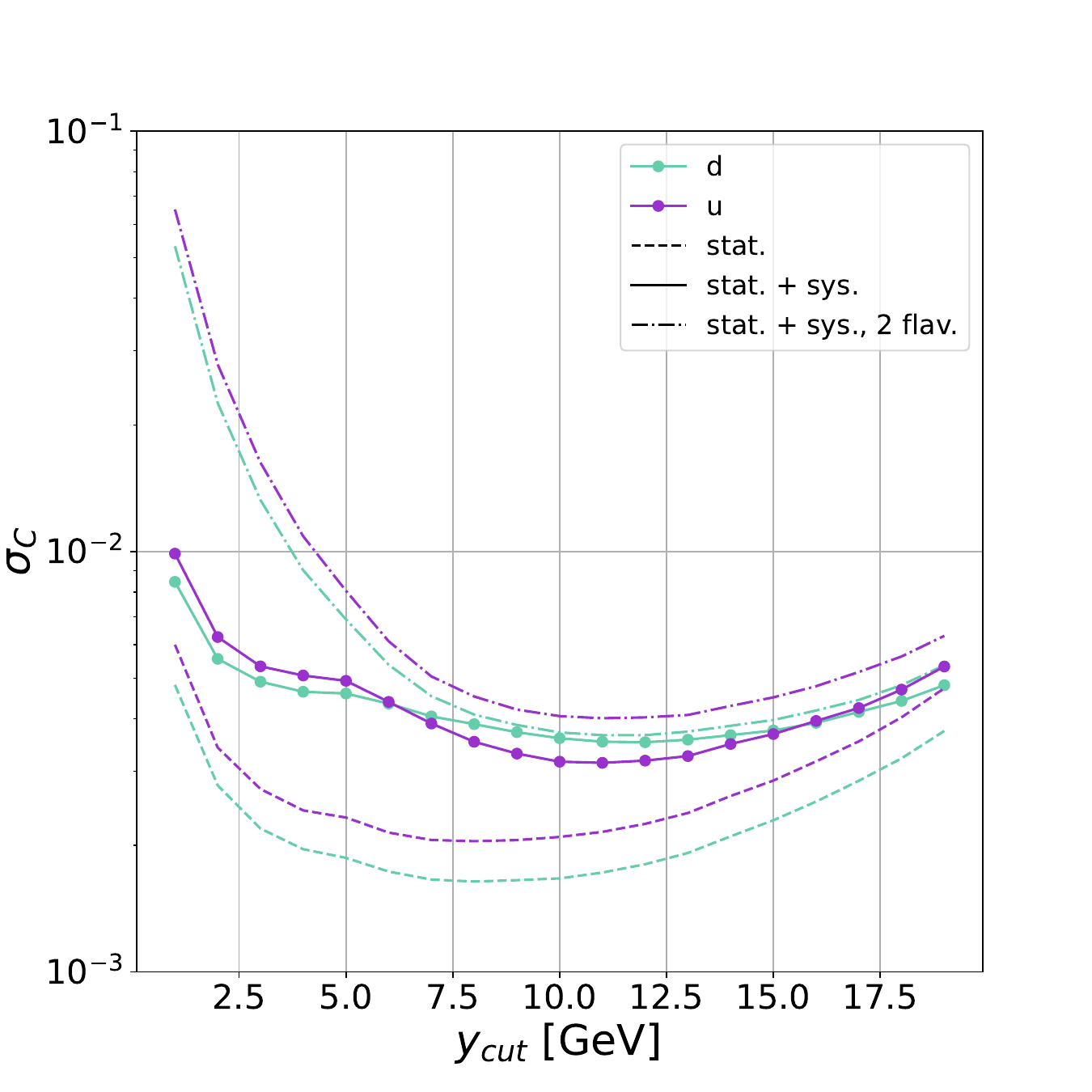}
    \includegraphics[width=0.46\textwidth]{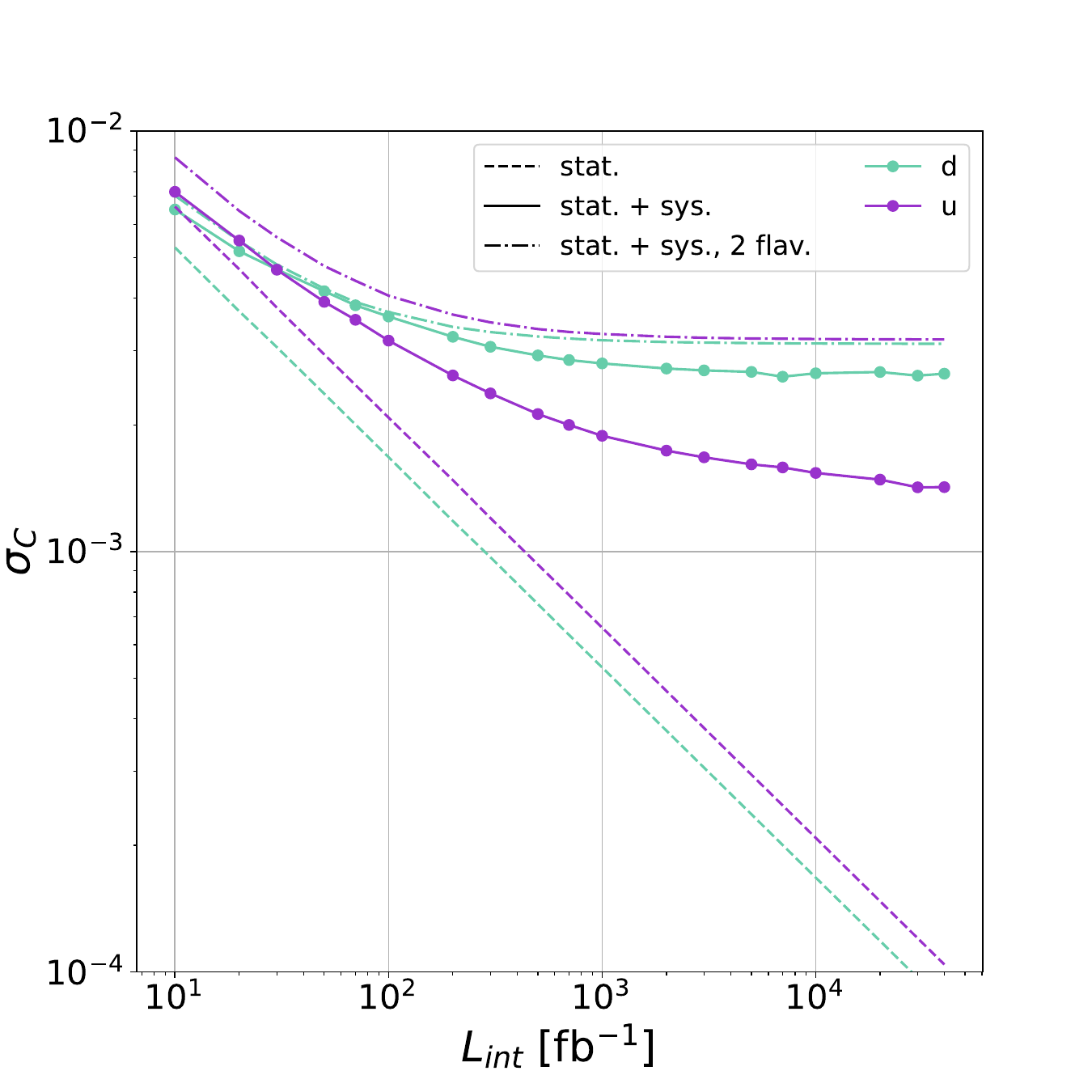}
    
    \caption{The uncertainty of the $d$ (aquamarine) and $u$ (purple) coupling measurement as a function of $y_{cut}$ (left) and the integrated luminosity (right). Dashed lines indicate statistical uncertainty only, solid lines with points -- statistical and systematic uncertainties for the 5-flavour fit, and dash-dotted lines -- statistical and systematic uncertainties for the 2-flavour fit. We assume 0.1\% uncertainty for all the contributions, except for the luminosity uncertainty (fixed at 10$^{-4}$). For the left plot, the luminosity was fixed at 100 fb$^{-1}$ and for the right plot, the value of $y_{cut}$ was set to 10 GeV.}
    \label{fig:2vs5flav}
\end{figure}

\subsection{5-flavour fit}

To improve our results, we decided to consider the entire data set including all jet flavours. This approach also incorporates events with heavy flavours (e.g. pairs of $b$ jets) and mixed events (e.g. one light jet and one $s$ jet). We hoped to improve the fitting by considering simultaneously all information available, which can constrain some uncertainties directly from the data. 

For this purpose, our statistical framework was extended to include all possible jet combinations, and the fitting procedure was repeated. A comparison between the 2- and 5-flavour fits is given in Figure \ref{fig:2vs5flav}. The improvement is especially remarkable for lower values of $y_{cut}$ and large luminosities. As expected, the inclusion of other channels allows for constraining systematic uncertainties from the data, which results in scaling of the measurement uncertainty with growing luminosity. It is also interesting to note that the precision for the $d$ quark becomes worse than for the $u$ quark at larger statistics which can be explained by the signal contamination originating from the $s$-quark channel.

\begin{figure}
    \centering
    \includegraphics[width=0.46\textwidth]{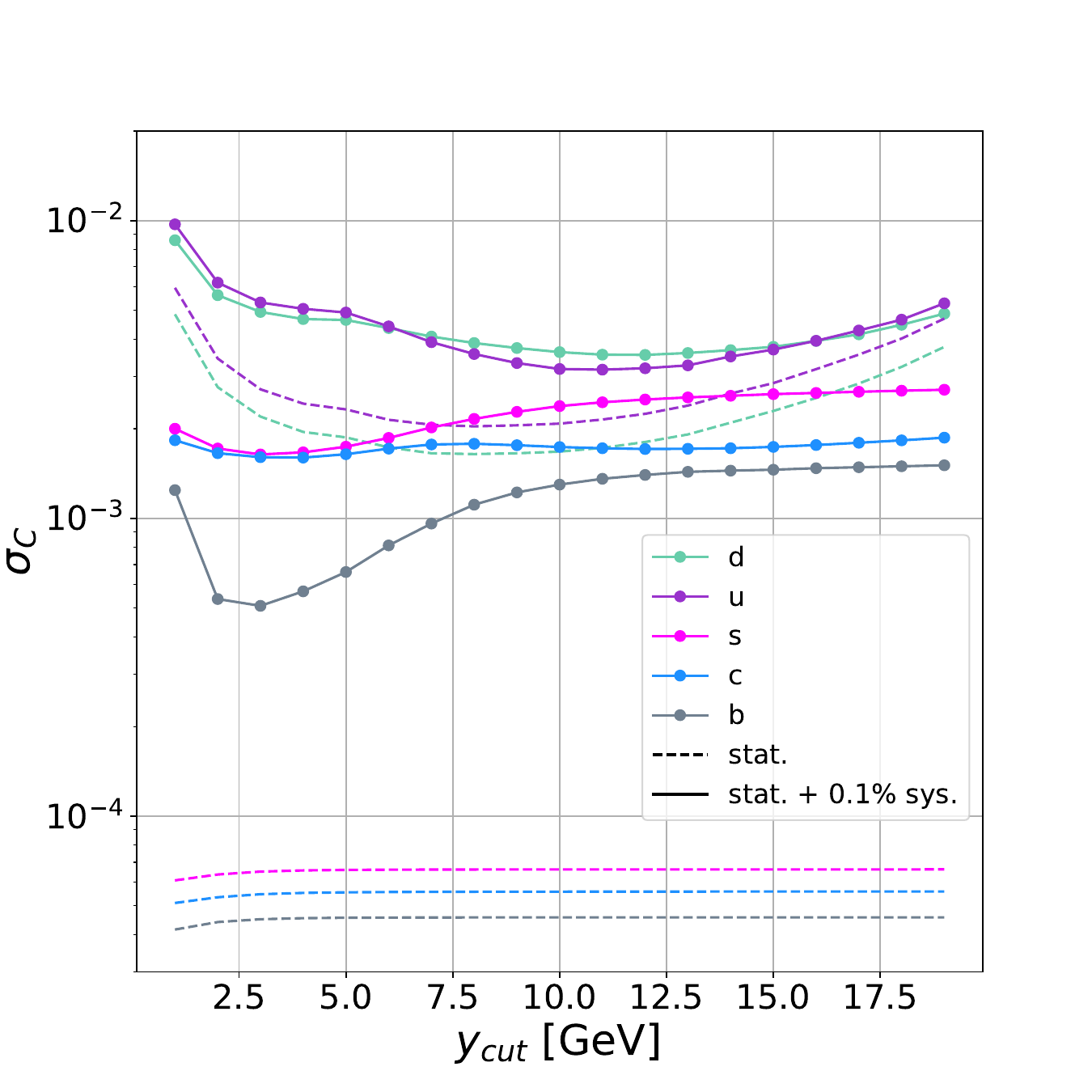}
    \includegraphics[width=0.46\textwidth]{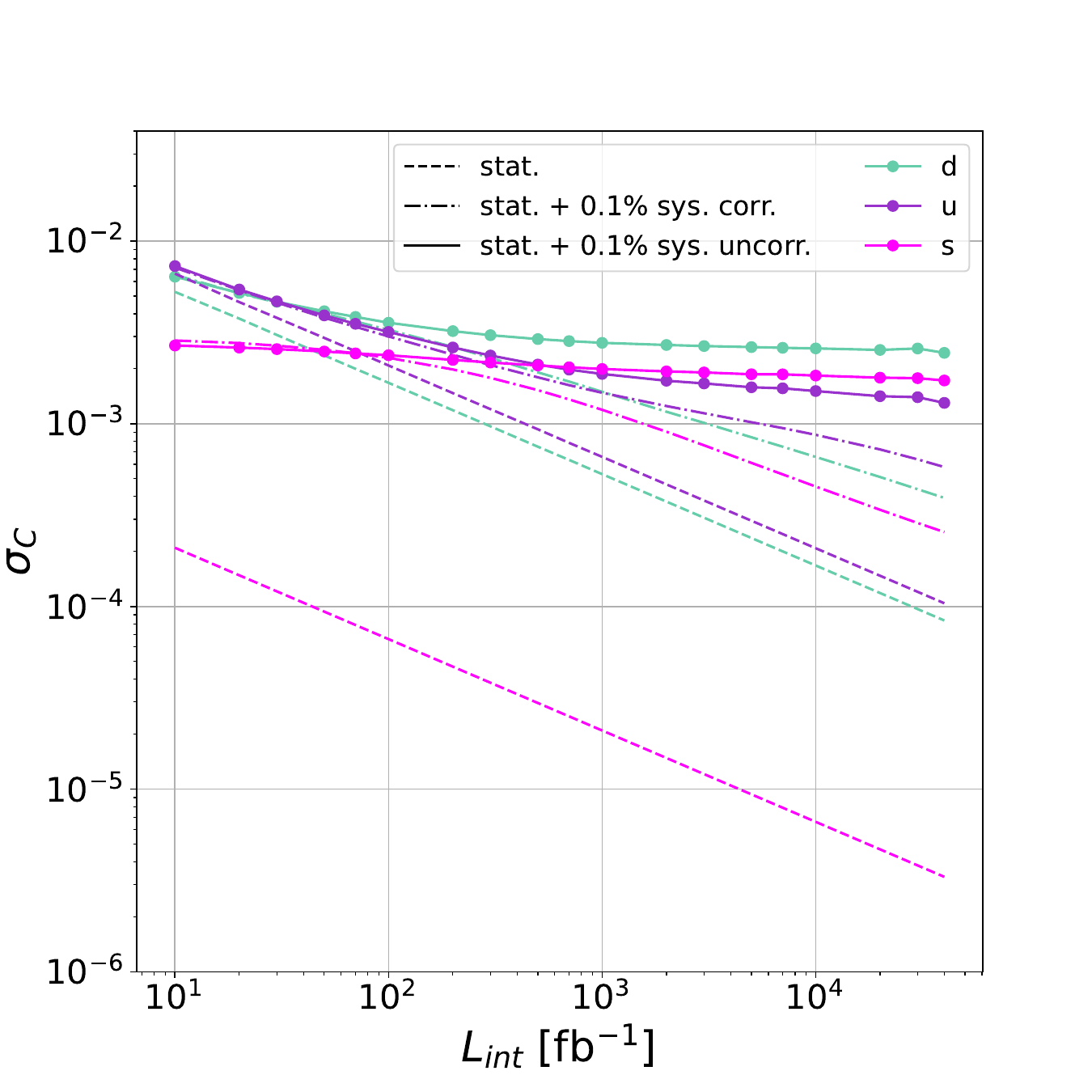}
    
    \caption{The uncertainty of the $d$ (aquamarine), $u$ (purple), $s$ (pink), $c$ (blue) and $b$ (gray) coupling measurement as a function of $y_{cut}$ (left) and the integrated luminosity (right). Dashed lines indicate statistical uncertainty only, the two other lines -- statistical and systematic uncertainties combined for two cases: solid lines with points for $u$- and $d$-tagging uncertainties uncorrelated, dash-dotted lines for $u$- and $d$-tagging uncertainties fully correlated. We assume 0.1\% uncertainty for all the contributions, except for the luminosity (fixed at 10$^{-4}$). For the left plot, the luminosity was fixed at 100 fb$^{-1}$ and for the right plot, the value of $y_{cut}$ was set to 10 GeV. Data for heavy quarks (correlated $u$- and $d$-tagging uncertainties) are skipped in the right (left) plot to improve readability.}
    \label{fig:5flav}
\end{figure}

Figure \ref{fig:5flav} gives more insight into the results obtained. In the left plot, we show the dependence of the results on $y_{cut}$ for all five flavours. Our statistical framework allows for setting stringent limits for all the quark flavours. In the most favourable case of the $b$ quark, our analysis developed for the light-quark case, can still provide results by a factor of four better than the current measurement. The light-quark couplings could be constrained at the sub-percent level, yielding an essential prospect for the electroweak precision measurements. It is also worth noting that the results for light quarks are significantly worse already at the statistical level, which is connected to the number of events tagged into this category. 

In the right plot, we present our results as a function of the integrated luminosity for the three lightest quarks. 
Additionally, we present lines showing the case in which tagging (and mis-tagging) probabilities are assumed to be identical for the two light quarks (their systematic variations are fully correlated), reducing the number of systematic uncertainties considered in the fit. For all the other results presented in this work, we followed the conservative path of assuming that all systematic variations are independent (they are fully uncorrelated).
The figure shows that this assumption has a crucial impact on the results. We emphasize that even though the $u$ and $d$ jets are classified into the same category of ``light jet'', the uncertainties for the two physical particles are in principle independent.

\subsection{Discussion of systematic uncertainties}

\begin{figure}
    \centering
    \includegraphics[width=0.32\textwidth]{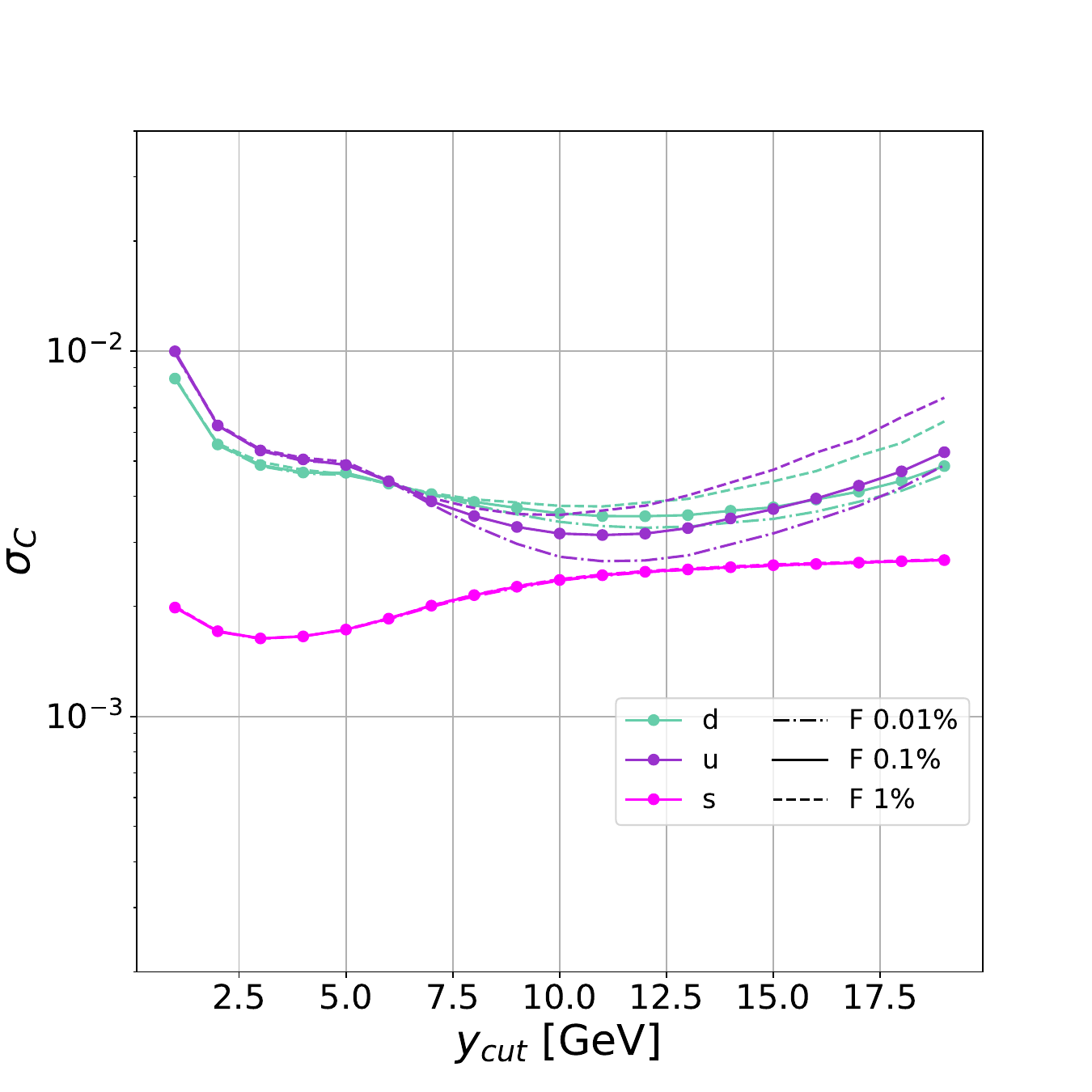}
    \includegraphics[width=0.32\textwidth]{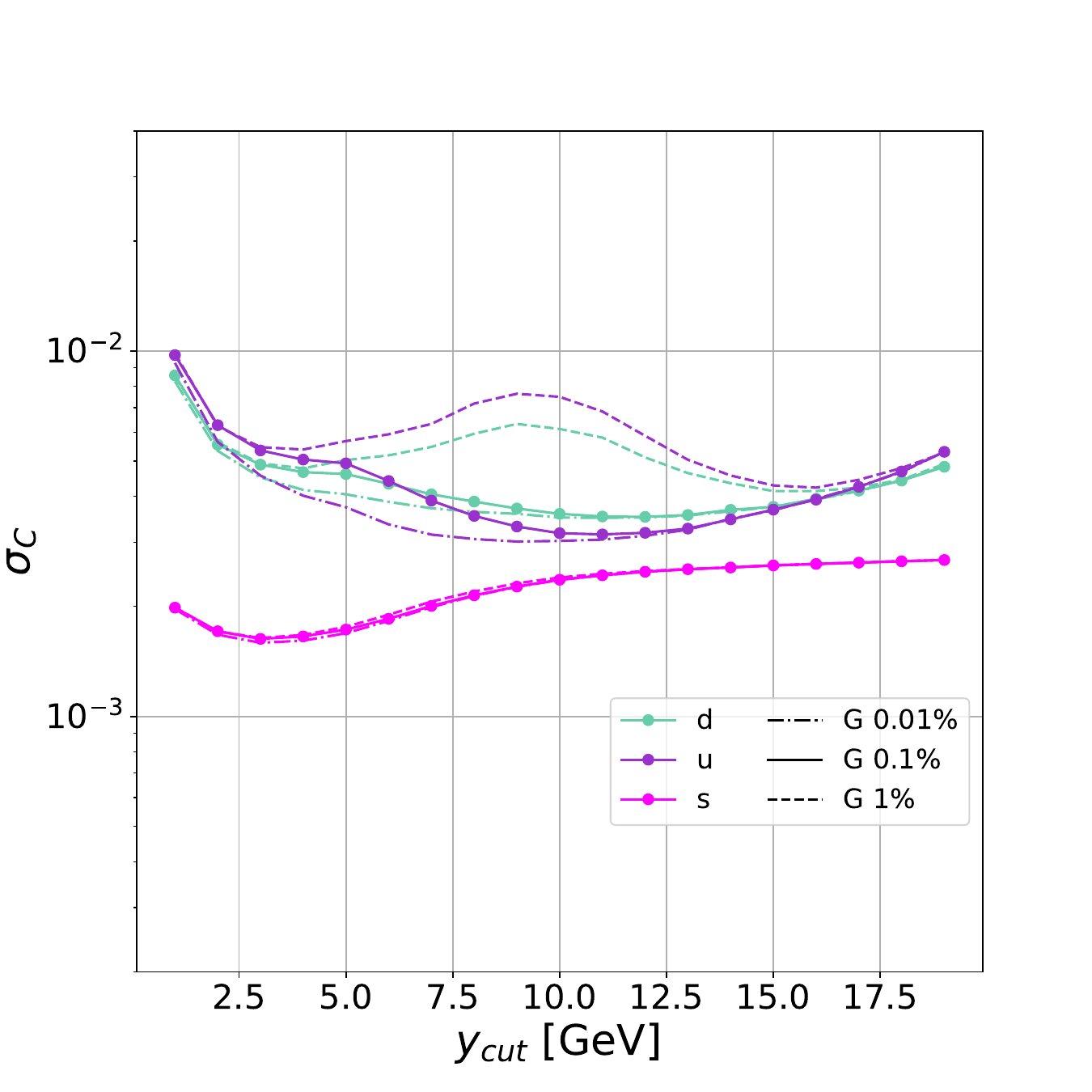}
    \includegraphics[width=0.32\textwidth]{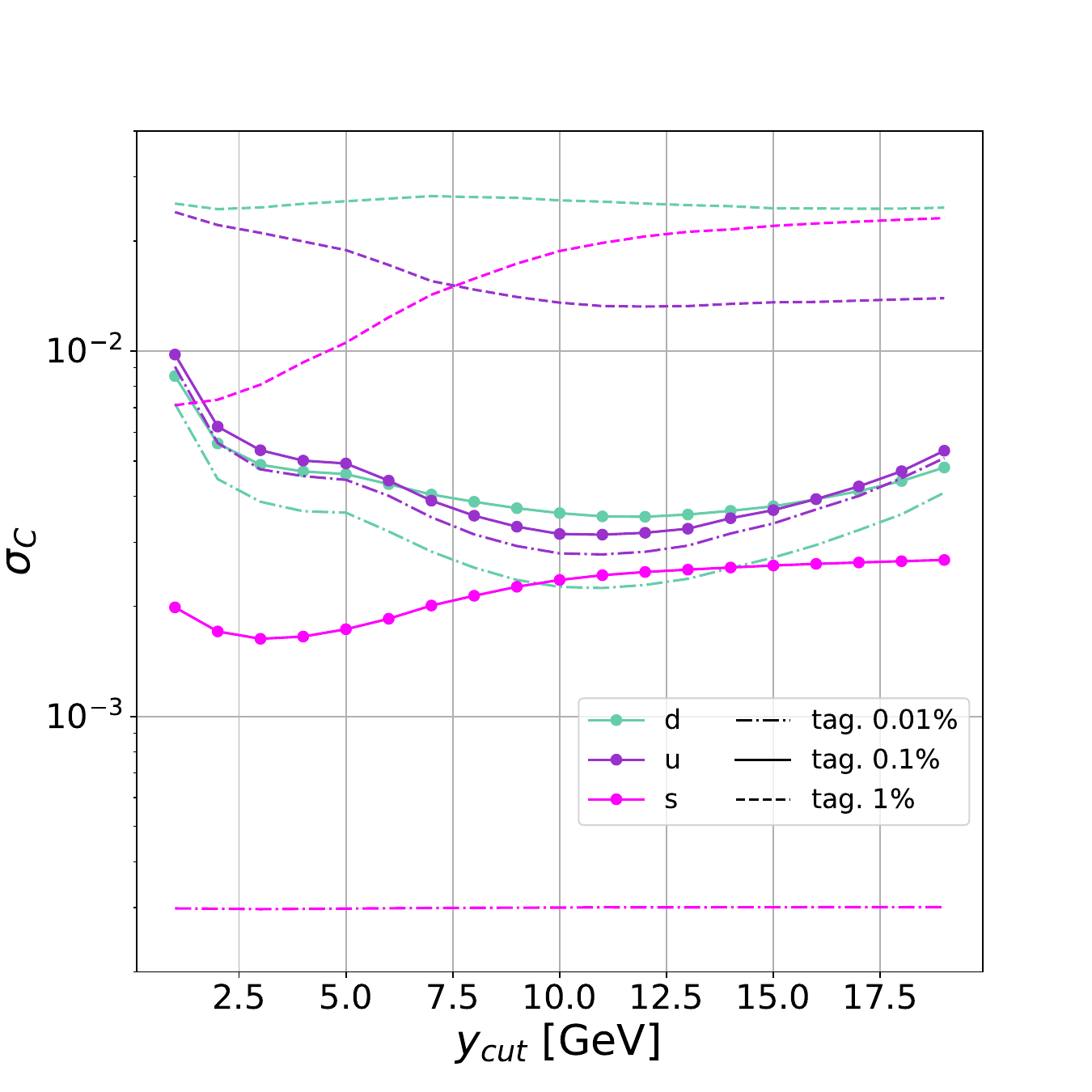}
    
    \caption{The uncertainty of the $d$ (aquamarine), $u$ (purple) and $s$ (pink) coupling measurement as a function of $y_{cut}$. The left plot shows the variation of the uncertainty of the acceptance of radiative events, the central one -- the uncertainty of the acceptance of non-radiative events, and the right one -- the tagging uncertainty. The solid lines indicate the uncertainty of 0.1\%, the dashed lines -- of 1\%, and the dash-dotted ones -- of 0.01\%. We assume 0.1\% uncertainty for all the unvaried contributions, except for the luminosity (fixed at 10$^{-4}$). The luminosity was set to 100 fb$^{-1}$.}
    \label{fig:5flav_unc}
\end{figure}

For the previous considerations, we set all the systematic uncertainties except for the luminosity uncertainty to the same value. However, it is illuminating to loosen this assumption and estimate the impact of particular sources of uncertainty. Figure \ref{fig:5flav_unc} shows this comparison by varying the uncertainties connected to selection efficiency of radiative events, hadronisation background and tagging, respectively. Three values are considered: 0.1\% (treated elsewhere as the ``reference'' value, and used also for non-varied sources of systematic uncertainties in each plot), 0.01\% and 1\%. The tagging uncertainty has a crucial impact on the results, while the dependence on the uncertainty of the acceptance of hadronic events is rather small and appears only for larger values of $y_{cut}$. The impact of the uncertainty on the acceptance of radiative events is marginal in the considered range. We stress that to fully explore the physics potential of future $Z$-pole runs at lepton colliders, a deep understanding of jet tagging is mandatory.

\begin{figure}
    \centering
    \includegraphics[width=0.52\textwidth]{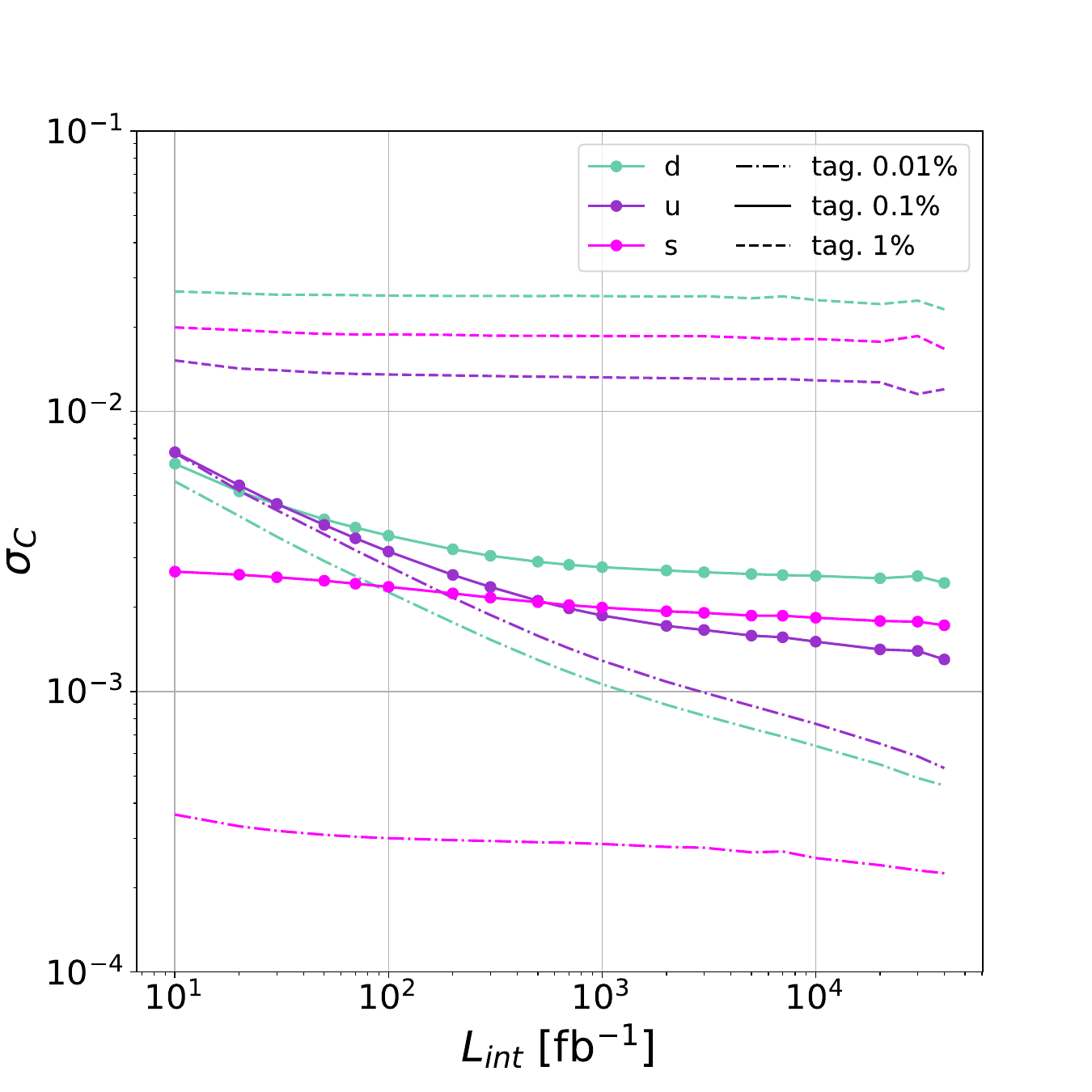}
    
    \caption{The uncertainty of the $d$ (aquamarine), $u$ (purple) and $s$ (pink) coupling measurement as a function of the integrated luminosity. The plot shows the variation of the tagging uncertainty; the solid lines indicate the uncertainty of 0.1\%, the dashed lines -- of 1\%, and the dash-dotted one -- of 0.01\%. We assume 0.1\% uncertainty for all the unvaried contributions, except for the luminosity (fixed at 10$^{-4}$). The $y_{cut}$ parameter was set to 10 GeV.}
    \label{fig:5flav_unc_lumi}
\end{figure}

In Figure \ref{fig:5flav_unc_lumi}, we show the results as a function of the luminosity for different values of the tagging uncertainty. The plot clearly shows that to benefit from higher data statistics, sub-permille tagging uncertainty must be achieved. According to our study, sub-percent level precision for the light-quark couplings to the $Z$ boson can be achieved at a future experiment collecting 100 fb$^{-1}$ of data, provided that the tagging uncertainty is reduced below 0.3\%, and the relative uncertainty on the background to the radiative sample from hadronisation photons remains at 1\% or below. The effect of the relative uncertainty on the radiative event selection efficiency is found to be marginal. With higher integrated luminosities, these requirements can be relaxed. For instance, at 40 ab$^{-1}$, the tagging uncertainty may be allowed up to 0.4\%, while the impact of both the radiative event selection efficiency uncertainty and the hadronisation photon background uncertainty becomes negligible.

\subsection{Discussion of correlations in the measurement}
\begin{figure}
    \centering
    \includegraphics[width=0.52\textwidth]{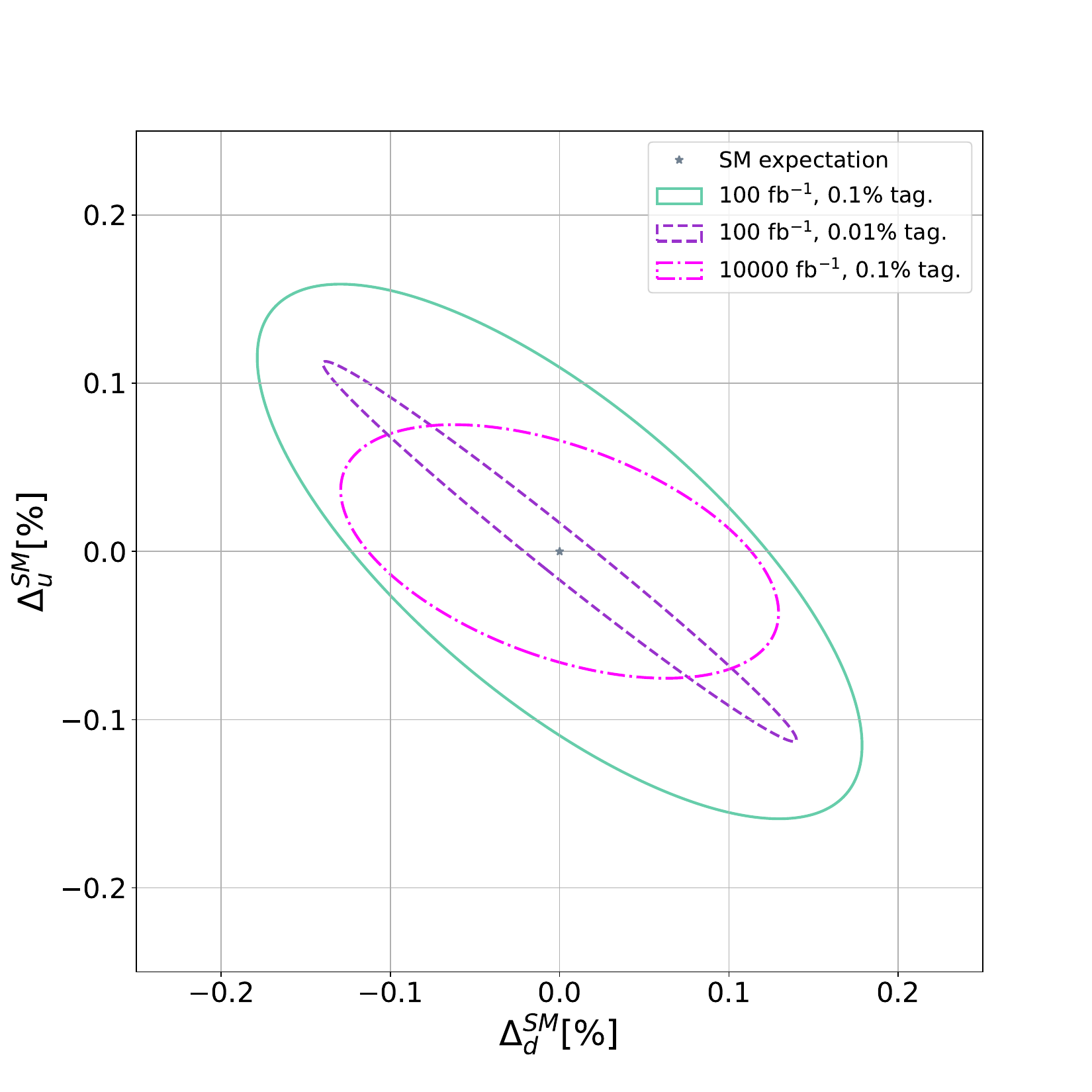}
    
    \caption{The correlation ellipses for the measurement of the $d$ and $u$ couplings for three different cases: 100 fb$^{-1}$ of collected data with tagging uncertainty of 0.1\% (solid aquamarine), 100 fb$^{-1}$ of collected data with tagging uncertainty of 0.01\% (dashed purple), 10000 fb$^{-1}$ of collected data with tagging uncertainty of 0.1\% (dash-dotted pink), respectively. The deviation from the SM value is shown. We assume 0.1\% uncertainty for all the other contributions, except for the luminosity (fixed at 10$^{-4}$). The $y_{cut}$ parameter was set to 10 GeV.}
    \label{fig:eli}
\end{figure}

The measurement aims to disentangle the dependence of the cross sections on the quark couplings, but the measured parameters of the SM are all correlated. By collecting more data of better quality, one can minimise this effect. In Figure \ref{fig:eli}, we show the correlation ellipses for the $d$ and $u$ quark couplings as their deviation from the SM values. We compare the improvement one can achieve by increasing luminosity or improving tagging efficiency. Both an improvement in the amount of data collected and a higher tagging efficiency change the sensitivity of this measurement and can significantly enhance the precision.

\section{Conclusions and outlook}
\label{sec:conclusions}
Future $e^+e^-$ colliders operating at the $Z$-pole will further constrain the Standard Model parameters and possibly pave the path beyond. Precision measurements of the $Z$-boson couplings to fermions will be possible thanks to very high data statistics. In this study, we established a dedicated event generation procedure including photons coming from different sources -- ISR, FSR, hadronisation and hadron decays. We simulated the detector response and studied photon isolation criteria. Finally, we estimated the expected uncertainties of the measurement of $u$ and $d$ quark couplings to the $Z$ boson, including statistical and systematic contributions. Our study shows that in order to profit from the extremely high luminosities foreseen for future $e^+e^-$ colliders, systematic uncertainties of the tagging efficiency should be brought down to the sub-permille level. High precision should also be secured for theoretical calculations and Monte Carlo simulations affecting the uncertainty of measuring radiative events and misidentifying hadronisation photons. For uncertainties of 0.1\%, one can expect a precision of about 0.3\% for the measurement of the light-quark couplings to the $Z$ boson, improving upon the statistically limited LEP data by at least an order of magnitude. In general, for integrated luminosities considered for future experiments at $e^+e^-$ colliders, the tagging uncertainty should achieve a precision of 0.3-0.4\% to measure the light-quark couplings to the $Z$ bosons at the sub-percent precision. For an integrated luminosity of 100\,fb$^{-1}$, the uncertainty on misidentifying hadronisation photons should be constrained at the level of 1\%, while its impact at an integrated luminosity of 40\,fb$^{-1}$ becomes negligible.
At the same time, our generic framework allows for constraining the precision of the $b$ ($c$) coupling up to 0.05\% (0.1\%), which yields an improvement of a factor of 5-20 and can potentially be further optimised. Additional insight could be gained by extending the analysis to higher-energy runs of the future $e^+e^-$ colliders, which has been deferred to future investigations. In order to further disentangle the different vector and axial vector contribution to the Z boson couplings of light quarks polarised beams would be very helpful, though some information could already be gathered by using forward-backward asymmetries at the $Z$-pole. Such a study is beyond the scope of this work. Moreover, a similar approach of distinguishing between $u$- and $d$-type quarks via photon radiation might be used to enhance other precision measurements in hadronic channels, including $H \to q\bar{q}$ decays.

\acknowledgments

The authors thank the ILD collaboration members for fruitful discussions, valuable comments and suggestions. The work was supported by the National Science Centre (Poland) under OPUS research project no.~2021/43/B/ST2/01778 and the Deutsche
Forschungsgemeinschaft (Germany) under
Germany's Excellence Strategy-EXC 2121 ``Quantum Universe''-390833306. Furthermore, we acknowledge support from
the COMETA COST Action CA22130,   by EAJADE - Europe-
America-Japan Accelerator Development and Exchange Programme (101086276), and from the International Center for Elementary Particle Physics (ICEPP), the University of Tokyo. 

\appendix
\section{Coupling fit framework}
\label{sec:app_stat}
In this Appendix, we review the details of the statistical analysis used for data fitting.

\subsection{Fit with statistical uncertainties only}
\label{s:data_fitting}
The cross section for each quark flavour can be presented as a sum of processes with and without hard photon radiation (the zero-photon exclusive and one-photon inclusive sample) in the final state:
\begin{equation}
    \sigma_q  =   \sigma_{0q}  + \sigma_{\gamma q},
\end{equation}
where the separation depends on the cuts defined for the photon isolation at the Matrix Element level and in the matching procedure.
The resulting fraction of hard radiative events at the generator level is then given by
\begin{equation}
    f_{q} = \frac{\sigma_{\gamma q}}{\sigma_q}.
\end{equation}

The definition of a \textit{radiative} event is imminently connected to photon reconstruction criteria. Thus, one has to define a set of cuts $y$ (including transverse momentum, isolation angle, etc.) describing reconstructable hard photons. 
Then, the measured radiative cross section depends on the factors $F_y \equiv F(y)$ and $G_y \equiv G(y)$ (assumed to be
quark-flavour independent for simplicity) which describe the experimental acceptance of radiative and non-radiative events (for which measurable photons come mostly from hadronisation and decays), respectively:
\begin{eqnarray}
    \sigma^{(y)}_{\gamma h} & = & \sum_{q} 
  F_y \cdot \sigma_{\gamma q} + G_y \cdot \sigma_{0 q} \\
  & = &  \sum_{q} \sigma_q \cdot \left( f_q F_y + (1-f_q) G_y \right).
\end{eqnarray}

Experimentally, based on the dedicated flavour-tagging algorithms, one can classify hadronic 2-jet-like events into $N$ flavour-based categories. One can classify events assuming same-flavour quark pair production, for example, $b\bar{b}$,
$c\bar{c}$ and lighter-quark final states ($N=3$) or complete flavour decomposition ($b\bar{b}$, $c\bar{c}$, $s\bar{s}$, $u\bar{u}$, $d\bar{d}$, $N=5$). On the other hand, one can also
classify events based on single jet-tagging results (which can differ
between the two jets in an event), resulting in more possible event categories (up to $N=15$ for full flavour decomposition). The advantage of the second approach is that the systematic uncertainties due to flavour-tagging efficiencies can be better constrained from data. 

Let $j$ be an index referring to a given flavour category, $j=1,\dots, N$. The total number of hadronic events expected in each category, after the
final selection and classification, can be written as:
\begin{eqnarray}
    N_j = \sum_q {\cal E}_q \cdot M_{q\,j} \cdot {\cal L}_{int} \cdot \sigma_q,
\end{eqnarray}
where ${\cal E}_q$ is the event selection efficiency for flavour $q$, with results obtained from the simulation in our study are provided in Table \ref{tab:eff}, ${\cal L}_{int}$ is the total integrated luminosity and 
$M_{q\,j}$ gives the probability of the  $q\bar{q}$ event being classified as category $j$. A similar equation can be set for radiative events:
\begin{eqnarray}
    N^{(y)}_{\gamma j} = \sum_q \Big(
 {\cal E}_{\gamma q} \cdot  M_{\gamma q\,j} \cdot f_q \cdot F_y  + 
 {\cal E}_{0 q} \cdot  M_{0 q\,j} \cdot (1-f_q) \cdot G_y 
 \Big) \cdot {\cal L}_{int} \cdot \sigma_q,
\end{eqnarray}
where ${\cal E}_{\gamma q}$ and ${\cal E}_{0 q}$ are the selection
efficiencies for radiative events and non-radiative background, i.e. ``fake'' radiative or category migration
events (not taking into account the photon selection cut efficiencies included in $F_y$ and
$G_y$, respectively; Table \ref{tab:eff}), $M_{\gamma q\,j}$ and $M_{0 q\,j}$ are the
jet-flavour classification matrices for radiative and non-radiative events. In the following, we will assume that $M_{q\,j} = M_{\gamma q\,j} = M_{0 q\,j}$. The following tagging-efficiency matrix was adapted from \cite{Liang:2023wpt}:
\begin{equation}
\begin{bmatrix}
0.773 & 0.190 & 0.031 & 0.006\\
0.773 & 0.190 & 0.031 & 0.006\\
0.311 & 0.645 & 0.038 & 0.006\\
0.107 & 0.068 & 0.795 & 0.030\\
0.026 & 0.007 & 0.059 & 0.908
\end{bmatrix}.
\end{equation}
The five rows of the matrix correspond to five quark flavours ($d$, $u$, $s$, $c$, $b$), and the four columns to four available jet classification labels (``light jet'', ``$s$ jet'', ``$c$ jet'', ``$b$ jet'').
To simplify the formula, we define two matrices:
\begin{eqnarray}
    A_{q\,j} =  {\cal E}_q \cdot M_{q\,j} \cdot {\cal L}_{int},
\end{eqnarray}
\begin{eqnarray}
    B_{q\,j} =   {\cal E}_{\gamma q} \cdot  M_{q\,j} \cdot {\cal L}_{int}
  \cdot f_q \cdot F_y   + 
        {\cal E}_{0 q} \cdot  M_{q\,j} \cdot {\cal L}_{int}
        \cdot (1-f_q) \cdot G_y
\end{eqnarray}
and we rewrite\footnote{We will treat the dependence on $y$ as implicitly assumed and drop the superscript in $N^{(y)}_{\gamma j}$.}:
\begin{eqnarray}
    N_{j} = \sum_q A_{q\,j}\cdot \sigma_q, \label{eq:nj}
\end{eqnarray}
\begin{eqnarray}
    N_{\gamma j} =  \sum_q B_{q\,j}\cdot \sigma_q.  \label{eq:ngj}
\end{eqnarray}

\begin{table}[]
\centering
\begin{tabular}{c|c|c|c|c|c}
$q$           & $d$     & $u$     & $s$     & $c$     & $b$     \\ \hline
${\cal E}_{q}$        & 0.955 & 0.955 & 0.947 & 0.949 & 0.936 \\ \hline
${\cal E}_{0 q}$     & 0.902 & 0.903 & 0.852 & 0.892 & 0.870 \\ \hline
${\cal E}_{\gamma q}$ & 0.917 & 0.922 & 0.876 & 0.914 & 0.888
\end{tabular}
\caption{Event selection efficiencies obtained from the simulation. See text for details.}
\label{tab:eff}
\end{table}

Collider experiments are expected to measure event numbers in $2N$ categories
(total hadronic and radiative) and we want to extract $M=5$ quark-level cross
sections which are directly related to the quark couplings. Assuming the number of events is large for each category $j$, we consider the $\chi^2$ function: 
\begin{eqnarray}
    \chi^2 = \sum_j \frac{(n_j - N_j)^2}{n_j} + 
   \sum_j \frac{(n_{\gamma j} - N_{\gamma j})^2}{n_{\gamma j}},
\end{eqnarray}
where $n_j$ and $n_{\gamma j}$ are the numbers of hadronic and radiative events, respectively, observed in given categories. 

Following the maximum likelihood principle, we calculate partial derivatives with respect to $\sigma_q$ and obtain a system of $M$ equations for cross section values, which can be represented in a matrix form:
\begin{eqnarray}
     \mathbb{H} \cdot \vec{\sigma} & = &  \mathbb{V}
\end{eqnarray}
or alternatively ($i = 1, \ldots, M$):
\begin{eqnarray}
     \sum_q \mathbb{H}_{iq} \cdot \sigma_q & = &  \mathbb{V}_{i},
\end{eqnarray}
where the Hessian matrix $\mathbb{H}$ and vector $\mathbb{V}$ are defined as:
\begin{eqnarray}
     \mathbb{H}_{iq} &=&  \frac{1}{2}
  \frac{\partial^2\chi^2}{\partial \sigma_i \partial \sigma_q} 
   =   \sum_j \frac{A_{ij} A_{qj}}{N_j} + 
  \sum_j \frac{B_{ij} B_{qj}}{N_{\gamma j}}, \\
  \mathbb{V}_i &=&  \sum_j \frac{n_j}{N_j} \, A_{ij} + \sum_j  \frac{n_{\gamma j}}{N_{\gamma j}} \, B_{ij} \;,
\end{eqnarray}
respectively. The solution can be found by inverting matrix $\mathbb{H}$:
\begin{eqnarray}
     \vec{\sigma} & = & \mathbb{H}^{-1} \cdot \mathbb{V} \\[2mm]
  \sigma_i & = & \sum_q  \left(\mathbb{H}^{-1}\right)_{iq} \;   \mathbb{V}_{q}.
\end{eqnarray}
The inverse of the Hessian matrix is also the covariance matrix for the
extracted cross sections. In particular, cross-section uncertainties
are given by the square root of the diagonal elements of the inverse matrix:
\begin{eqnarray}
     \Delta \sigma_i & = &\sqrt{ \left(\mathbb{H}^{-1}\right)_{ii}}.
\end{eqnarray}
Since the cross section linearly depends on the quark coupling, and the proportionality constant can be precisely determined from theoretical calculations, we assume that the uncertainty on the coupling is the same as the uncertainty of the cross-section measurement, $\Delta \sigma_i \equiv \sigma_C$.

\subsection{Including systematic uncertainties}
\label{s:systematic_uncertainties}
The matrices $A_{q\,j}$ and $B_{q\,j}$ can be obtained from Monte Carlo
simulations of the physics processes and detector performance with
negligible statistical uncertainty. However, they are a subject of
multiple systematic uncertainties which have to be taken into account.
Assuming $K$ independent systematic uncertainties described by
variations $\delta_k$, we can extend the $\chi^2$ formula to include dependence on the systematic variations: 
\begin{eqnarray}
     \chi^2(\vec{\delta}) =
  \sum_j \frac{(n_j - N_j(\vec{\delta}))^2}{n_j}  +  \sum_j \frac{(n_{\gamma j} -N_{\gamma j}(\vec{\delta}))^2}{n_{\gamma j}} + \sum_k \delta_k^2
     + \sum_q 2 \lambda_q \; w_q(\vec{\delta})\; ,
\end{eqnarray}
where the third term corresponds to the assumed normal distribution for the systematic variations $\delta_k$ and the last term includes Lagrange multipliers $\lambda_q$ introduced to enforce proper normalisation of the tagging probabilities (the sum of their variations, $w_q$, has to be zero for each quark flavour $q$).  The dependence of the numbers of expected hadronic and radiative events, $N_j$ and $N_{\gamma j}$, on the considered systematic effects is described by the variation of matrices $A_{qj}$ and $B_{qj}$ (see eqs~\ref{eq:nj} and \ref{eq:ngj}). To allow for a semi-analytical solution, we restrict ourselves to the linear dependence of $A_{q\, j}$ and $B_{q\,j}$ on the systematic variations, $\delta_k$, and rewrite:
\begin{eqnarray}
     A_{q\, j} = A_{q\, j}^{MC} + \sum_k a^k_{q\, j} \cdot \delta_k,\\
  B_{q\, j} = B_{q\, j}^{MC} + \sum_k b^k_{q\, j} \cdot \delta_k,
\end{eqnarray}
where $a^k_{q\, j}$ and $b^k_{q\, j}$ correspond to $1 \, \sigma$ variations of $A_{q\, j}^{MC}$ and $B_{q\,j}^{MC}$ (which are the nominal values obtained from the Monte Carlo simulation) due to systematic uncertainty $k$.
In the linear approximation, it is also useful to rewrite the experimentally measured cross sections by defining the expected cross section (theoretical value calculated within the SM), $\sigma^{th}_q$, and the cross-section deviation parameters, $\Delta_q$:
\begin{eqnarray}
     \sigma_q = \sigma^{th}_q + \Delta_q.
\end{eqnarray}
When solving the maximum likelihood problem with systematic uncertainties included,
one tries to find a solution for $\Delta_q$, $\delta_k$ (and $\lambda_q$) assuming small deviations from the nominal
predictions, $|\Delta_q| \ll \sigma_q^{th}$, $|a_{qj}^k| \ll A_{qj}$ and $|b_{qj}^k| \ll B_{qj}$, 
so that terms including products
of deviations  ($\Delta_q \Delta_{q'}$, $\delta_k \delta_{k'}$ and $\Delta_q \delta_k$) can be neglected. The problem can be reduced to solving the system of $2M+K$ linear equations, similar to the one previously discussed:
\begin{eqnarray}
     \tilde{\mathbb{H}} \cdot
  \left(\! \begin{array}{c} \vec{\Delta}\\ \vec{\delta} \\ \vec{\lambda} \end{array} \!\right)
  =  \tilde{\mathbb{V}}.
\end{eqnarray}

\bibliographystyle{JHEP}
\bibliography{bibliography.bib}

\end{document}